\documentclass[twocolumn,english,superscriptaddress,citeautoscript,showpacs,preprintnumbers,amsmath,amssymb,prl,floatfix,footinbib]{revtex4}
\usepackage[utf8x]{inputenc}
\usepackage[scaled=2]{helvet}
\usepackage[T1]{fontenc}
\usepackage[utf8]{luainputenc}
\usepackage{multirow}
\usepackage{textcomp}
\usepackage{amsmath}
\usepackage{amssymb}
\usepackage{graphicx}
\usepackage{subscript}
\usepackage{epstopdf}
\makeatletter

\newcommand{\lyxmathsym}[1]{\ifmmode\begingroup\def\b@ld{bold}
  \text{\ifx\math@version\b@ld\bfseries\fi#1}\endgroup\else#1\fi}

\@ifundefined{textcolor}{}
{%
 \definecolor{BLACK}{gray}{0}
 \definecolor{WHITE}{gray}{1}
 \definecolor{RED}{rgb}{1,0,0}
 \definecolor{GREEN}{rgb}{0,1,0}
 \definecolor{BLUE}{rgb}{0,0,1}
 \definecolor{CYAN}{cmyk}{1,0,0,0}
 \definecolor{MAGENTA}{cmyk}{0,1,0,0}
 \definecolor{YELLOW}{cmyk}{0,0,1,0}
}

\usepackage[scaled=2]{helvet}\usepackage{amstext}

\makeatletter
\@ifundefined{textcolor}{}{\definecolor{BLACK}{gray}{0}\definecolor{WHITE}{gray}{1}\definecolor{RED}{rgb}{1,0,0}\definecolor{GREEN}{rgb}{0,1,0}\definecolor{BLUE}{rgb}{0,0,1}\definecolor{CYAN}{cmyk}{1,0,0,0}\definecolor{MAGENTA}{cmyk}{0,1,0,0}\definecolor{YELLOW}{cmyk}{0,0,1,0}}

\usepackage{dcolumn}
\usepackage{bm}
\usepackage{times}
\usepackage{hyperref}
\hypersetup{colorlinks=true,linkcolor=red,citecolor=blue}
\makeatother

\usepackage{babel}

\makeatother
\usepackage{babel}

\begin{document}


\title{Coexistence of Eu-antiferromagnetism and pressure-induced superconductivity
\\in EuFe$_2$As$_2$ single crystal}

\author{W. T. Jin}
\affiliation{Jülich Centre for Neutron Science JCNS at Heinz Maier-Leibnitz Zentrum (MLZ), Forschungszentrum Jülich GmbH, Lichtenbergstraße 1, D-85747 Garching, Germany}

\author{Y. Xiao}
\email[y.xiao@pku.edu.cn]{}
\affiliation{School of Advanced Materials, Peking University, Shenzhen Graduate School, Shenzhen 518055, China}

\author{S. Nandi}
\affiliation{Jülich Centre for Neutron Science JCNS and Peter Grünberg Institut PGI, JARA-FIT, Forschungszentrum Jülich GmbH, D-52425 Jülich, Germany}

\author{S. Price}
\affiliation{Jülich Centre for Neutron Science JCNS at Heinz Maier-Leibnitz Zentrum (MLZ), Forschungszentrum Jülich GmbH, Lichtenbergstraße 1, D-85747 Garching, Germany}

\author{Y. Su}
\affiliation{Jülich Centre for Neutron Science JCNS at Heinz Maier-Leibnitz Zentrum (MLZ), Forschungszentrum Jülich GmbH, Lichtenbergstraße 1, D-85747 Garching, Germany}

\author{K. Schmalzl}
\affiliation{Jülich Centre for Neutron Science JCNS at Institut Laue-Langevin (ILL), Forschungszentrum Jülich GmbH, Boite Postale 156, 38042 Grenoble Cedex 9, France}

\author{W. Schmidt}
\affiliation{Jülich Centre for Neutron Science JCNS at Institut Laue-Langevin (ILL), Forschungszentrum Jülich GmbH, Boite Postale 156, 38042 Grenoble Cedex 9, France}

\author{T. Chatterji}
\affiliation{Institut Laue-Langevin, 6 rue Jules Horowitz, 38042 Grenoble Cedex 9, France}

\author{A. Thamizhavel}
\affiliation{Department of Condensed Matter Physics and Material Sciences, Tata Institute of Fundamental Research, Homi Bhabha Road, Colaba, Mumbai 400 005, India}

\author{Th. Br\"{u}ckel}
\affiliation{Jülich Centre for Neutron Science JCNS and Peter Grünberg Institut PGI, JARA-FIT, Forschungszentrum Jülich GmbH, D-52425 Jülich, Germany}
\affiliation{Jülich Centre for Neutron Science JCNS at Heinz Maier-Leibnitz Zentrum (MLZ), Forschungszentrum Jülich GmbH, Lichtenbergstraße 1, D-85747 Garching, Germany}
\affiliation{Jülich Centre for Neutron Science JCNS at Institut Laue-Langevin (ILL), Forschungszentrum Jülich GmbH, Boite Postale 156, 38042 Grenoble Cedex 9, France}

\date{\today}

\begin{abstract}

By performing high-pressure single-crystal neutron diffraction measurements, the evolution of structure and magnetic ordering in EuFe$_2$As$_2$ under hydrostatic pressure were investigated. Both the tetragonal-to-orthorhombic structural transition and the Fe spin-density-wave (SDW) transition are gradually suppressed and become decoupled with increasing pressure. The antiferromagnetic order of the Eu sublattice is, however, robust against the applied pressure up to 24.7 kbar, without showing any change of the ordering temperature. Under the pressure of 24.7 kbar, the lattice parameters of EuFe$_2$As$_2$ display clear anomalies at 27(3) K, well consistent with the superconducting transition observed in previous high-pressure resistivity measurements. Such an anomalous thermal expansion around \emph{T}$_c$ strongly suggests the appearance of bulk superconductivity and strong electron-lattice coupling in EuFe$_2$As$_2$ induced by the hydrostatic pressure. The coexistence of long-range ordered Eu-antiferromagnetism and pressure-induced superconductivity is quite rare in the EuFe$_2$As$_2$-based iron pnictides.

\end{abstract}

\pacs{74.70.Xa, 61.05.fm, 74.62.Fj, 74.62.-c}
\maketitle

\section{Introduction}

Large efforts have been undertaken to explore the knowledge of superconductivity (SC) in high \emph{T}$_c$ Fe-based superconductors since they were discovered in 2008 \cite{Kamihara, Johnston}. General phase diagrams of \emph{A}Fe$_2$As$_2$-type ("122" with \emph{A} = Ba, Ca, and Eu, etc.) iron pnictides clearly show that the parent compounds undergo a tetragonal-to-orthorhombic structural phase transition accompanied with the formation of antiferromagnetically ordered spin density wave (SDW) state. The static magnetic order of parent compounds can be suppressed and SC emerges concomitantly with appropriate charge carrier doping or the application of external pressure. The complexity of phase diagrams in charge carrier doped systems revealed the intimate relationship between structural, magnetic and superconducting transitions. In contrast to charge carrier doping, the application of external pressure is considered as a more straightforward and cleaner way to induce SC, since there are no disorder effects caused by chemical inhomogeneity. Previous studies on "122" iron pnictide superconductors have shown that the application of high-pressure can not only induce SC but also tune the structural/magnetic phases \cite{Chu,Sefat,Kimber, Kreyssig,Goldman,Prokes, Wu,Tomic}. Therefore, it is worthwhile to investigate in details the entanglement between the structure, magnetism and SC in pressure-induced "122" iron pnictide superconductors by neutron diffraction, although it is very challenging from an experimental point of view.

Among various parent compounds of "122" iron pnictide superconductors, EuFe$_2$As$_2$ is considered as an interesting member given the fact that the \emph{A} site is occupied by Eu$^{2+}$, an \emph{S}-state rare-earth ion possessing a 4$\mathit{f}$$^{7}$ electronic configuration with the electron spin \emph{S} = 7/2 \cite{Zapf_2017}. Two successive magnetic phase transitions have been identified at 190 and 19 K, corresponding to SDW ordering of the itinerant Fe moments and A-type antiferromagnetic ordering of the localized Eu$^{2+}$ moments, respectively \cite{Raffius, Jeevan_08, Xiao_09}. The SDW transition in EuFe$_2$As$_2$ can be suppressed continuously by applying external pressure due to the weakening of Fe-Fe exchange interactions. Furthermore, SC with \emph{T}$_c$ $\sim$ 30 K can be induced in EuFe$_2$As$_2$ by external pressure, which appears in a narrow pressure region in the vicinity of the critical pressure \emph{P}$_c$ $\sim$ 25 kbar \cite{Miclea, Terashima, Kurita}. Upon the application of pressure up to 80 kbar, a pressure-induced magnetic transition of the Eu$^{2+}$ moment from antiferromagnetic to ferromagnetic ordering was suggested and a collapsed tetragonal phase was found for EuFe$_2$As$_2$ \cite{Matsubayashi, Uhoya}.

Although many studies about the pressure effects on EuFe$_2$As$_2$ have been performed during the past few years, as of yet no account of the lattice response in pressure-induced superconducting phase has been reported. Furthermore, the magnetic order of Eu$^{2+}$ moment in the pressure-induce superconducting phase is still not clarified, although it was argued based on high-pressure transport and x-ray spectroscopy measurements that the antiferromagnetic order of Eu$^{2+}$ moment is relatively robust against the pressure and persist below 60 kbar \cite{Matsubayashi}. However, it was recently confirmed in Co-doped EuFe$_2$As$_2$ that the pressure-induced SC is also compatible with the ferromagnetic order of the Eu sublattice  \cite{Jin_17}. Therefore, it is interesting to study and clarify the interplays between the structure, magnetism, and SC under external pressure using an \emph{in situ} high-pressure neutron diffraction technique. In this report, hydrostatic-pressure neutron-diffraction experiments were performed on a EuFe$_2$As$_2$ single crystal to investigate the magnetism of the Eu sublattice in pressure-induced superconducting state, and the couplings between the lattice, Fe-SDW and SC.

\section{Experimental Methods}

High-pressure single-crystal neutron diffraction experiments were carried out on the thermal neutron two-axis diffractometer D23 at the Institut Laue-Langevin (Grenoble, France). The crystal mounted into the pressure cell was in shape of a platelet with approximate dimensions of 5~$\times$~3 $\times$~1~mm$^3$ with a total mass of 30 mg, which was cut from the same piece of crystal used for ambient-enviroment neutron diffraction as reported in Ref. \onlinecite{Xiao_09}. Due to the large neutron absorption cross-section of Eu, the incident neutron wavelength of 1.28 $\buildrel_\circ \over {\mathrm{A}}$ was selected for the measurement. To investigate the evolution of the structure and magnetic order of EuFe$_2$As$_2$ with hydrostatic pressure, a clamped pressure cell equipped with a cylinder-shaped sample holder was used. A mixture of ethanol and methanal was adopted as the pressure transmitting medium to guarantee the hydrostaticity of the applied pressure. The EuFe$_2$As$_2$ crystal was oriented with the \emph{a*-c*} scattering plane horizontally aligned, allowing the magnetic reflections from both the Fe and Eu sublattices as well as the nuclear reflections in the (\emph{H}0\emph{L}) reciprocal plane to be accessible. For convenience, we will use the orthorhombic notation throughout this paper unless otherwise stated. A small piece of NaCl crystal was also put into the pressure cell at the side of the EuFe$_2$As$_2$ crystal, with its \emph{ab} plane horizontally aligned. By tracking the (200) reflection of NaCl, the lattice constant of cubic NaCl can be accurately determined and used as a standard indicator of the applied pressure value, based on the well-established equation of state of NaCl \cite{Decker}. After mounting the sample and NaCl into the clamped cell, the whole pressure cell was then mounted into a standard cryostat for single crystal neutron diffraction measurements between 2 and 300 K.

\section{Results and Discussions}

\subsection{\label{sec:level2} Pressure-induced decoupling between Fe-SDW and structural transitions}

The splitting of the tetragonal (\emph{H}\emph{H}0)$_T$ nuclear reflections with decreasing temperature can be considered as the indication for a tetragonal-to-orthorhombic structural phase transition in "122" iron pnictide superconductors. After the application of hydrostatic pressure on the EuFe$_2$As$_2$ single crystal, Q-scans were performed accross the tetragonal (220)$_T$ Bragg reflection as a function of temperature. Under the pressure of \emph{P} = 24.7 kbar, a single peak was observed at 172 K for the tetragonal (220)$_T$ reflection and it split into orthorhombic (400)$_O$ and (040)$_O$ reflections when temperature reached 95 K, as shown in the insets of Fig. 1(a). By tracking the evolution of (400)$_O$ and (040)$_O$ peak positions, the structural transition temperatures of EuFe$_2$As$_2$ at 11.5 and 24.7 kbar are determined to be 180(2) and 165(2) K, respectively. Given that the same structural transition of EuFe$_2$As$_2$ in ambient environment takes place at 190 K \cite{Xiao_09}, it is clear that the structural transition temperature of EuFe$_2$As$_2$ decreases gradually with increasing applied pressure.

\begin{figure}[!htb]
\includegraphics[width=8.5cm,height=11.6cm]{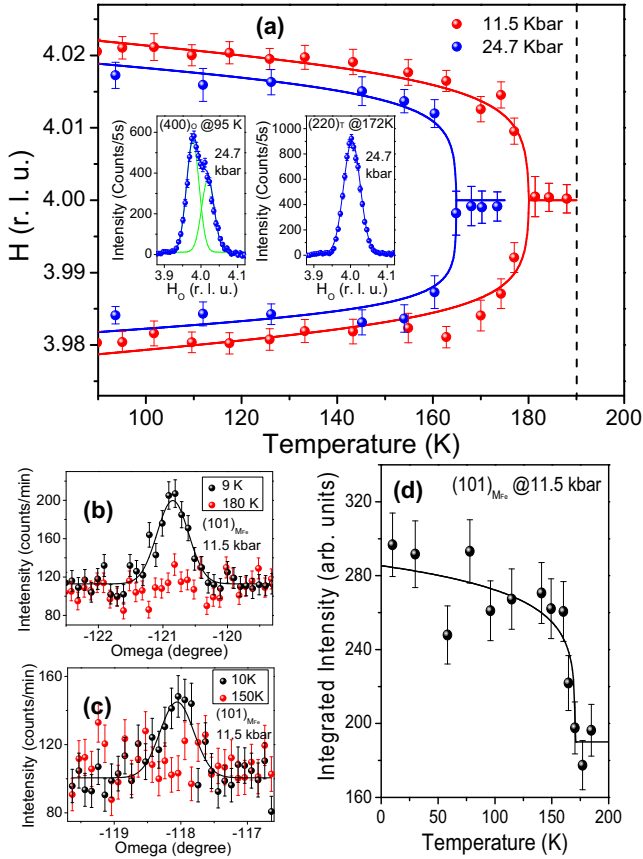}
\caption{\label{fig:epsart} (Color online) (a) Variation of peak positions of orthorhombic (400)$_O$ and (040)$_O$ reflections under different applied pressures. The insets show the scans across orthorhombic (400)/(040)$_O$ reflections at 95 K and tetragonal (220)$_T$ reflection at 172 K, respectively, along the orthorhombic (100) direction or tetragonal (110) direction, under the pressure of \emph{P} = 24.7 kbar. (b) The (101)$_M$ magnetic reflection measured at 9 K and 180 K under \emph{P} = 11.5 kbar. (c) The (101)$_M$ magnetic reflection measured at 10 K and 150 K under \emph{P} = 24.7 kbar. The solid lines in (a), (b) and (c) represent fittings to the peaks using a Gaussian profile. (d) Temperature dependence of the integrated intensity of (101)$_M$ magnetic Bragg reflection under \emph{P} = 11.5 kbar. }
\end{figure}

Apart from the nuclear Bragg reflections, the magnetic reflections originated from long-range SDW ordering of the Fe sublattice were also followed at low temperature under pressure. As shown in Fig. 1(b) and (c), weak signals were observed for (101)$_M$ reflection under \emph{P} = 11.5 kbar, while they were still present but largely suppressed under \emph{P} = 24.7 kbar. The Fe-SDW ordering temperature under \emph{P} = 11.5 kbar is estimated to be 170(5) K according to the temperature dependence of the integrated intensity of the (101)$_M$ reflection (See Fig. 1(d)). The ordering temperature under \emph{P} = 24.7 kbar can not be determined due to the weakness of magnetic intensity, but it is definitely below 150 K as indicated by Fig. 1(c).

Our neutron diffraction measurements of the order parameters clearly demonstrated the decoupling between the structural and Fe-SDW transitions in EuFe$_2$As$_2$ upon the application of hydrostatic pressure. It is worth noting that the separation between these two transitions is a well-confirmed feature in doped BaFe$_2$As$_2$ and EuFe$_2$As$_2$ \cite{Nandi_10, Lu_13, Jin_16, Jin_PhaseDiagram}. Therefore, hydrostatic pressure seems to play a similar role to the chemical doping in tuning the structural and Fe-SDW transitions in EuFe$_2$As$_2$. In addition, it was observed in a high-pressure electrical transport measurement that the cusp in the resistivity of EuFe$_2$As$_2$ single crystal continuously shifted to lower temperature with increasing pressure, ascribing to the structural or Fe-SDW transitions \cite{Kurita}. Compared with the temperature scales of our neutron data, it is indicated that the cusp observed in the resistivity curve is corresponding to the Fe-SDW transition.

\subsection{\label{sec:level2} Robust AFM order of Eu$^{2+}$ moments}

In contrast to the largely suppressed SDW ordering in the Fe sublattice with applied pressure, up to \emph{P} = 24.7 kbar, which is close to the maximum pressure value a clamped cell can exert in reality, the antiferromagnetic order of the Eu$^{2+}$ moments was found to be quite robust against the applied pressure. Q-scans at 2 K under \emph{P} = 24.7 kbar, as shown in Fig. 2(a) and (b), revealed the appearance of magnetic reflections from the Eu sublattice at the forbidden Bragg peak positions with the propagation vector of \emph{k} = (001). This suggests an antiferromagnetic interlayer coupling between adjacent Eu layers, which is the same as the case under ambient pressure \cite{Xiao_09}. The magnetic origin of these reflections is  evidenced from their disppearance at 20 K (see Fig. 2(c)). From the temperature dependencies of the integrated intensities of both (003)$_M$ and (203)$_M$, the antiferromagnetic ordering temperature of the Eu sublattice  can be determined to be 19.0(5) K under \emph{P} = 24.7 kbar, which is unchanged compared with the value measured in ambient enviroment \cite{Xiao_09}.

In other words, applying a hydrostatic pressure up to \emph{P} = 24.7 kbar does not provide a large enough perturbation to EuFe$_2$As$_2$ for changing the magnetic order of the Eu sublattice. This is consisent with the pressure-temperature phase diagram of EuFe$_2$As$_2$ constructed based on high-pressure transport and x-ray spectroscopy measurements  \cite{Matsubayashi}, in which the proposed transition of the magnetic state of the Eu sublattice from antiferromagnetic to ferromagnetic (AFM-FM) only takes place under a much higher hydrostatic pressure (\emph{P} > 60 kbar) and the $\mathit{T_{Eu}}(P)$ curve is rather flat for \emph{P} < 25 kbar.

\begin{figure}
\includegraphics[width=8.5cm,height=6.5cm]{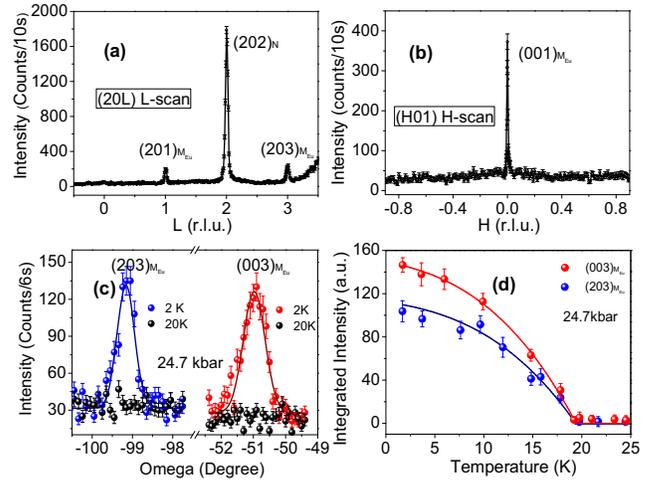}
\caption{\label{fig:epsart} (Color online) (a) L-scan along the (2, 0, L) direction and (b) H-scan along the (H, 0, 1) direction at 2 K under \emph{P} = 24.7 kbar. (c) Rocking scan of (203)$_M$ and (003)$_M$ reflections at 2 and 20 K, in which the solid lines represent fittings to the peaks using a Gaussian profile. (d) The temperature dependencies of the integrated intensity of (003) and (203) reflections under \emph{P} = 24.7 kbar.}
\end{figure}

\subsection{\label{sec:level2} Response of the lattice to pressure-induced superconductivity}

The lattice instability in high-temperature superconductors have been extensively studied as it may provide evidences of electron-lattice interactions. It was observed that the lattice parameters show some anomalies at the superconducting transition temperature \emph{T}$_c$ for cuprates and magnesium diboride \cite{Fujii, Jorgensen}. According to the pressure-temperature phase diagram of EuFe$_2$As$_2$ \cite{Kurita}, the pressure-induced superconducting phase is confined in a narrow pressure region between 24 and 31 kbar. Under \emph{P} = 24.7 kbar, we have deduced the lattice parameters of EuFe$_2$As$_2$ from the \emph{H}-scans across orthorhombic (400)/(040)$_O$ and \emph{L}-scans across orthorhombic (008)$_O$ reflections at different temperatures in the low-temperature region. As presented in Fig. 3, all lattice parameters (\emph{a}, \emph{b}, and \emph{c}) exhibit clear anomalies at 27(3) K, which is well consistent with the superconducting transition temperature \emph{T}$_c$ = 29(2) K observed in high-pressure resistivity measurements \cite{Kurita}. This anomalous thermal expansion around \emph{T}$_c$ strongly suggests the appearance of bulk superconductivity and strong electron-lattice coupling in EuFe$_2$As$_2$ induced by the application of hydrostatic pressure of \emph{P} = 24.7 kbar.

\begin{figure}[!htb]
\includegraphics[width=8.5cm,height=8.8cm]{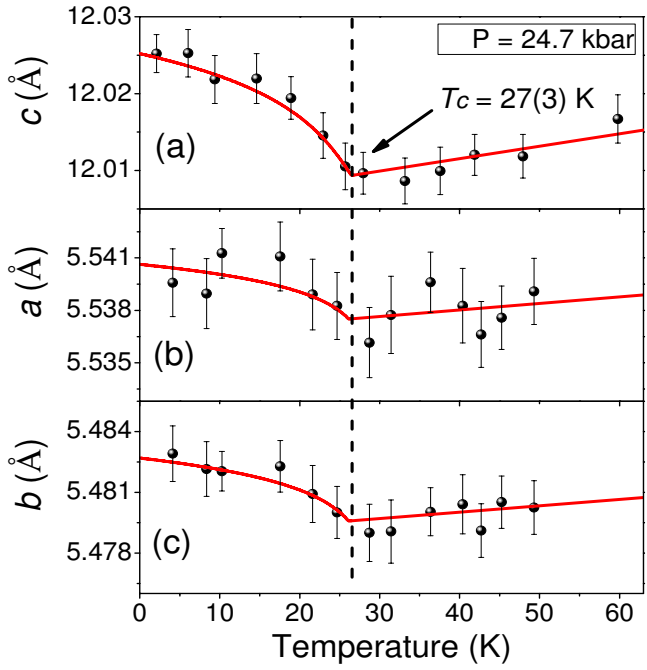}
\caption{\label{fig:epsart} (Color online) (a) The temperature dependencies of the lattice parameters \emph{a}, \emph{b}, and \emph{c} under \emph{P} = 24.7 kbar. The solid lines are guides to the eye.}
\end{figure}

The responses of the crystal lattice to superconductivity have been observed in many \emph{A}Fe$_2$As$_2$-type superconductors. By using hig-resolution dilatometry, Hardy \emph{et al.} observed clearly the anisotropic lattice responses (along \emph{c} and \emph{a} direction, respectively) to the superconducting order in a nearly optimally-doped Ba(Fe$_{0.92}$Co$_{0.08}$)$_2$As$_2$ single crystal with a large uniaxial-pressure dependence of the critical temperature \emph{T}$_c$ \cite{Hardy}. It is worth noting that the anisotropic lattice response was deduced from the linear expansivities with uniaxial stress instead of hydrostatic pressure. The uniaxial pressure dependence along two crystallographic directions can be largely compensated under hydrostatic pressure and result in a smaller negative thermal expansion. In the present work, the hydrostatic pressure is \emph{in situ} applied during the neutron diffraction measurements. Hence, the small negative thermal expansions along all crystallographic axes reflect the intrinsic response of the lattice to superconductivity in EuFe$_2$As$_2$ compounds. The response can be attributed to the spontaneous strain generated in the superconducting phase via a strong electron-lattice interaction.

\subsection{\label{sec:level2} Pressure-temperature phase diagram of EuFe$_2$As$_2$}

\begin{figure}[!htb]
\centering{}\includegraphics[width=1\columnwidth]{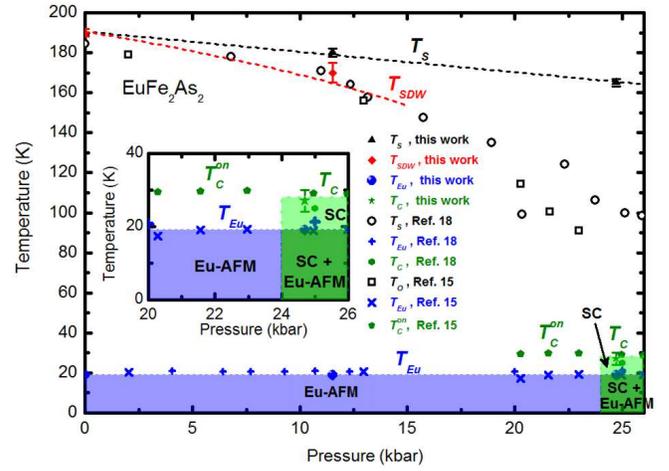}
\caption{\label{fig:epsart} (Color online) Pressure-temperature phase diagram of EuFe$_2$As$_2$ determined from the neutron diffraction measurements. The filled triangles, diamonds and spheres represent the structural transition temperatures, Fe-SDW ordering temperatures, and Eu-AFM ordering temperatures, respectively, determined from neutron diffraction measurements. The open circles and squares represent the temperatures of anomalies in the resistivity curves in Ref. \onlinecite{Matsubayashi} and \onlinecite{Miclea}, respectively.  The crossings represent the  Eu-AFM ordering temperatures determined from ac magnetic susceptibility and resistivity measurements in Ref. \onlinecite{Matsubayashi} and \onlinecite{Miclea}, respectively. The filled star marks the superconducting transition inferred from Fig. 3. The filled pentagons and hexagon represent the onset superconducting transition and zero-resistivity critical temperature evidenced by the resistivity measurements in Ref. \onlinecite{Matsubayashi} and \onlinecite{Miclea}, respectively. The inset is an enlarged view for the pressure region between 20 kbar and 26 kbar. The area marked by blue is the Eu-AFM phase, while that marked by green is the superconducting phase. The dotted line are guides to the eye. }
\end{figure}

Based on the results of our neutron diffraction study presented above, we have established a pressure-temperature phase diagram of EuFe$_2$As$_2$ in the low-pressure region (\emph{P} < 26 kbar) (Fig. 4). Compared with the data of transition temperatures extracted from the phase diagrams reported in Ref. \onlinecite{Miclea} and  \onlinecite{Matsubayashi}, the temperatures of anomalies in the resistivity curves (denoted as $\mathit{T_{S}}$ in Ref. \onlinecite{Matsubayashi} and $\mathit{T_{0}}$ in Ref. \onlinecite{Miclea}) line well with the Fe-SDW ordering temperatures  ($\mathit{T_{SDW}}$ = 170(5) K for P = 11.5 kbar and $\mathit{T_{SDW}}$ < 150 K for P = 24.7 kbar) determined from our neutron diffraction measurements. The decoupling between the structural and Fe-SDW transitions with increasing pressure was not reported in both references, as the electrical resistivity measurements can not differentiate between them. However, with the access to both nuclear and magnetic scattering, we were able to identify the structural and Fe-SDW transitions distinctively using neutron diffraction. The pressure-induced seperation between these two transitions resembles that observed in electron-doped EuFe$_2$As$_2$ \cite{Jin_16, Jin_PhaseDiagram}, suggesting a similar role of hydrostatic pressure to chemical doping in tuning the phase transitions. The antiferromagnetic order of the Eu$^{2+}$ moments, however, is quite robust against the applied pressure up to \emph{P} = 24.7 kbar, without showing any change of the ordering temperature $\mathit{T_{Eu}}$. To verify the scenario of the AFM-FM transition under higher pressure (\emph{P} > 60 kbar) as proposed in Ref. \onlinecite{Matsubayashi}, a Paris-Edinburgh pressure cell is needed to achieve a much higher hydrostatic pressure in further neutron diffraction experiments.

Inferred from the unambiguous responses of lattice parameters and the observation of zero-resistivity for the comparable value of applied pressure as reported in Ref. \onlinecite{Miclea}, \onlinecite{Kurita} and  \onlinecite{Matsubayashi}, EuFe$_2$As$_2$ is expected to enter a unique pressure-induced bulk superconducting state below \emph{T}$_c$ = 27(3) K. Therefore, as shown in Fig.4, there is a specific phase regime in which the long-range antiferromagnetic order of the Eu$^{2+}$ moments coexist with the pressure-induced SC. This is distinct from the well-documented doping-induced coexistence of Eu-ferromagnetism and the superconductivity \cite{Ren_09, Jin_13, Nandi_14, Nandi_14_neutron, Jin_15}, where a spontaneous vortex state is expected to account for the compromise between those two antagonistic phenomenon \cite{Jiao}. Such a coexistence of Eu-AFM and SC was reported for Eu$_{0.5}$K$_{0.5}$Fe$_2$As$_2$ with K doping \cite{Jeevan_08_K}, but the antiferromagnetism of the Eu sublattice was proposed to be of a short-range nature there. In our case, the Eu-AFM in the pressure-induced superconducting phase is clearly long-range ordered, making such a phase quite rare in the EuFe$_2$As$_2$-based iron pnictides.

\section{Conclusion}

In summary, by performing high-pressure single-crystal neutron diffraction measurements, the evolution of structure and magnetic ordering in EuFe$_2$As$_2$ with hydrostatic pressure were investigated. Both the structural phase transition and the Fe-SDW transition are gradually suppressed and become decoupled with increasing pressure. The antiferromagnetic order of the Eu sublattice is, however, robust against the applied pressure up to 24.7 kbar, without showing any change of the ordering temperature. Under the pressure of 24.7 kbar, the lattice parameters of EuFe$_2$As$_2$ display clear anomalies at 27(3) K, well consistent with the superconducting transition observed in previous high-pressure resistivity measurements. Such an anomalous thermal expansion around \emph{T}$_c$ strongly suggests the appearance of bulk superconductivity and strong electron-lattice coupling in EuFe$_2$As$_2$ induced by the hydrostatic pressure. The coexistence of long-range ordered Eu-antiferromagnetism and pressure-induced superconductivity is quite rare in the EuFe$_2$As$_2$-based iron pnictides.

\section{\label{sec:level1}Acknowledgment}

The authors are grateful to Jean-Luc Laborier and Claude Payre from Service of Advanced Neutron Environments (SANE) at Institut Laue-Langevin for providing assistance with high-pressure equipement during the neutron diffraction experiments.

\appendix

\end{document}